\documentclass[prd,preprint,floatfix,nofootinbib]{revtex4}
\usepackage{epsfig}
\usepackage{graphicx}
\usepackage{axodraw}
\usepackage{pstricks}

\def\lsim{\mathrel {\vcenter {\baselineskip 0pt \kern 0pt
    \hbox{$<$} \kern 0pt \hbox{$\sim$} }}}
\def\gsim{\mathrel {\vcenter {\baselineskip 0pt \kern 0pt
    \hbox{$>$} \kern 0pt \hbox{$\sim$} }}}
\def\slashchar#1{\setbox0=\hbox{$#1$}           
 \dimen0=\wd0                                 
  \setbox1=\hbox{/} \dimen1=\wd1               
\ifdim\dimen0>\dimen1                        
  \rlap{\hbox to \dimen0{\hfil/\hfil}}      
  #1                                        
  \else                                        
 \rlap{\hbox to \dimen1{\hfil$#1$\hfil}}   
   /                                         
  \fi}                                         %
\def\cpto{\mathrel {\vcenter {\baselineskip 0pt \kern 0pt
    \hbox{$CP$} \kern 0pt \hbox{$\longrightarrow$} }}}
\def\cptof{\mathrel {\vcenter {\baselineskip 0pt \kern 0pt
    \hbox{$~CP$} \kern 0pt \hbox{$\longleftrightarrow$} }}}

\begin{document}

\baselineskip=15pt
\preprint{}

\title{$T$-odd correlations from $CP$ violating anomalous top-quark couplings revisited}

\author{Oleg Antipin and G. Valencia}

\email{oaanti02@iastate.edu}
\email{valencia@iastate.edu}

\affiliation{Department of Physics, Iowa State University, Ames, IA 50011.}

\date{\today}

\vskip 1cm
\begin{abstract}

We revisit the effect of $CP$ violating anomalous top-quark couplings  in $t\bar{t}$ production and decay.  We consider $t\bar{t}$ production through gluon fusion (and light $q{\bar q}$ annihilation) followed by top-quark decay into $bW$ or $b\ell\nu$.  We find explicit analytic expressions for all the triple products generated by the anomalous couplings that fully incorporate all spin correlations. Our results serve as a starting point for numerical simulations for the LHC.

\end{abstract}

\pacs{PACS numbers: 12.15.Ji, 12.15.Mm, 12.60.Cn, 13.20.Eb,
13.20.He, 14.70.Pw}

\maketitle

\section{Introduction}

With the upcoming start of the LHC, it is timely to revisit the question of $CP$ violation searches in high energy observables. A particularly useful concept for these searches is that of inclusive observables such as those defined in terms of jet momenta in Ref.~\cite{Donoghue:1987ax}.  These permit the construction of `null-tests': that is,  observables that vanish in the limit of $CP$ conservation. Many examples have been studied for $p\bar{p}$ and $e^+e^-$ colliders in detail \cite{tprods}. These observables take the form of triple product correlations which we refer to as `$T$-odd' because they change sign under the reversal of direction of all three momenta and are not necessarily $CP$-odd \footnote{These observables are sometimes referred to in the literature as ``naive-$T$''-odd  to distinguish this ``$T$'' from time reversal invariance.}. 
 
At the LHC, the $pp$ initial state is not a $CP$ eigenstate. Nevertheless, it is possible to construct a null test of $CP$ by focusing on a suitable final state. For example, in Ref.~\cite{Valencia:2005cx} we illustrated that this was possible by studying a triple product correlation in $H \to t\bar{t}\to b\bar{b}W^+W^-$ that originates in a spin correlation in $t\bar{t}$ production. In this case the reaction with the good $CP$ properties  is the Higgs decay, and the LHC is viewed as a Higgs factory. We then extended that idea to the same final state without the intermediate Higgs. For this purpose we considered the LHC to be a $t\bar{t}$ factory and then worked with this as the initial state. In particular at LHC energies, $t\bar{t}$ production via gluon fusion is dominant; and it is possible to construct the null-tests for the reaction $gg\to t {\bar t}$ (or to consider subsequent $t$ decay). The $gg$ initial state is not a $CP$ eigenstate, but summing over the gluon color and spin degrees of freedom can produce truly $CP$ odd observables much in the same manner as the jet observables of Ref.~\cite{Donoghue:1987ax}. This was illustrated in Ref.~\cite{Valencia:2005cx} using the simple example of a $Ht{\bar t}$ induced $CP$ violation. Of course, even in this case the remnants of the two initial protons are not a $CP$ eigenstate and could conceivably  fake a given $CP$ odd asymmetry. Although this type of background must eventually be studied, for now we assume that the $t{\bar t}$ state can be cleanly identified. 

In this paper we extend the study of Ref.~\cite{Valencia:2005cx} to consider the case of dimension 5,  $CP$-violating, anomalous top-quark couplings. This case describes in principle all models in which there are no new particles (beyond those already present in the SM) within reach of the LHC. It does not result in very large signals, but it illustrates the kind of observables that can be constructed for other models.

$CP$ violating anomalous $tbW$ couplings have been considered in the literature before. The ones we use here were originally defined in Ref.~\cite{Bernreuther:1992be}. Several aspects of these couplings have been studied in connection to hadron colliders \cite{Ma:1991ry,Brandenburg:1992be,Bernreuther:1993hq,hadronanom}, including a detailed numerical study of the ATLAS sensitivity \cite{Sjolin:2003ah}  and also in the context of $e^+e^-$ colliders Ref.~\cite{eeanom}.  The anomalous couplings have also been studied at length without special emphasis on $CP$ violating observables \cite{otheranom}. A recent numerical study of the ATLAS sensitivity to these couplings is Ref.~\cite{AguilarSaavedra:2007rs}.

The main new result from our calculation is a complete analytic expression for all the $T$-odd correlations in the process $gg \to t {\bar t} \to b {\bar b} \ell^+ \nu \ell^- {\bar \nu}$. These analytic expressions are relatively simple and fully incorporate all the spin correlations behind the observables. They are well suited for implementation in simulations that use the narrow width approximation for both the top-quark and $W$ propagators.

\section{Mixed helicity framework for $gg \to {t\bar t} \to b {\bar b} WW$.}

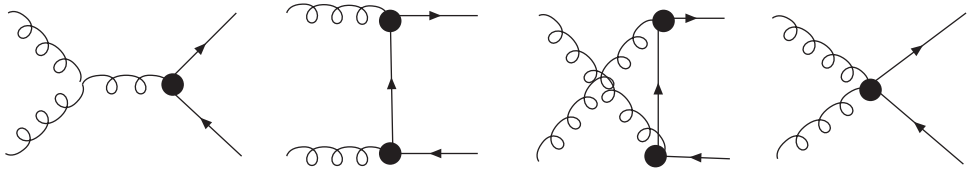
\begin{figure}[htb]
\begin{center}
\fcolorbox{white}{white}{
  \begin{picture}(362,61) (57,-53)
    \SetWidth{0.5}
    \SetColor{Black}
    \Gluon(59,5)(86,-21){3.5}{3.43}
    \Gluon(87,-22)(58,-48){3.5}{3.43}
    \Gluon(87,-22)(120,-22){3.5}{2.57}
    \ArrowLine(119,-21)(145,5)
    \ArrowLine(147,-49)(119,-23)
    \Vertex(121,-22){4.24}
    \Gluon(164,5)(202,4){3.5}{3.43}
    \Gluon(164,-49)(202,-49){3.5}{3.43}
    \ArrowLine(236,-48)(204,-48)
    \ArrowLine(204,-48)(203,5)
    \ArrowLine(204,4)(236,4)
    \Vertex(203,-48){4.24}
    \Vertex(203,2){4.24}
    \Gluon(347,3)(383,-23){3.5}{4.29}
    \Gluon(349,-50)(383,-24){3.5}{3.43}
    \ArrowLine(419,-52)(385,-23)
    \ArrowLine(384,-22)(420,5)
    \Vertex(384,-24){4.24}
    \Gluon(259,4)(306,-48){3.5}{6.86}
    \Vertex(306,2){4.24}
    \ArrowLine(331,-50)(305,-49)
    \ArrowLine(304,3)(329,3)
    \ArrowLine(304,-49)(304,2)
    \Vertex(303,-49){4.24}
    \Gluon(259,-51)(303,2){3.5}{6.86}
  \end{picture}
}
\end{center}
\caption{Diagrams responsible for $CP$ asymmetry in top-quark pair production via gluon fusion: $s$-channel, $t$-channel, $u$-channel and seagull.}
\label{fig:prod}
\end{figure}

\begin{figure}[htb]
\begin{center}
\fcolorbox{white}{white}{
  \begin{picture}(187,180) (96,-90)
    \SetWidth{0.5}
    \SetColor{Black}
    \GOval(156,1)(15,15)(0){0.882}
    \Gluon(102,-45)(146,-10){4.5}{5.14}
    \ArrowLine(166,13)(219,64)
    \ArrowLine(219,64)(266,90)
    \ArrowLine(263,-90)(220,-63)
    \ArrowLine(221,-63)(166,-11)
    \Photon(219,63)(248,27){4.5}{4}
    \Photon(222,-64)(248,-27){4.5}{4}
    \ArrowLine(247,26)(282,52)
    \ArrowLine(282,9)(248,26)
    \ArrowLine(246,-28)(284,-10)
    \ArrowLine(283,-55)(247,-27)
    \GBox(215,-67)(226,-59){0.882}
    \GBox(216,59)(227,67){0.882}
    \Gluon(97,43)(144,8){4.5}{5.14}
    \Text(178,-8)[lb]{\Large{\Black{$\Gamma_P$}}}
    \Text(198,68)[lb]{\Large{\Black{$\Gamma_D$}}}
    \Text(192,-80)[lb]{\Large{\Black{$\Gamma_{\bar{D}}$}}}
  \end{picture}
}
\end{center}
\caption{Decomposition of $t\bar{t}$ production and decay vertices with helicity amplitudes.}
\label{fig:fact}
\end{figure}
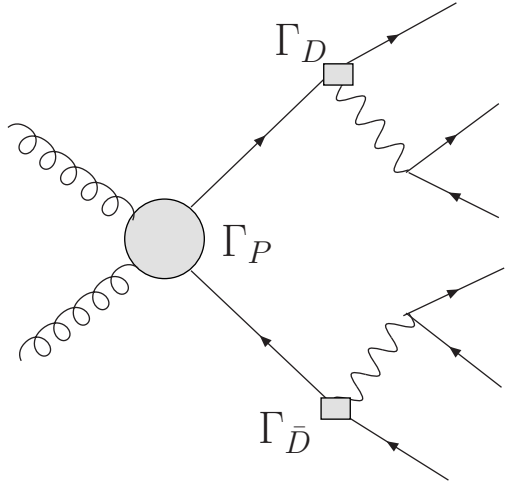

The dominant mechanism for production of $t\bar{t}$ pairs at the LHC is gluon fusion and we concentrate on it now.  For this source of $t\bar{t}$ pairs there are four relevant diagrams  shown in Figure~\ref{fig:prod} that we will consider. The first three diagrams are the usual $s,t,u$ channels in the SM. We will also consider the possibility of CP violation in the $ttg$ vertex as described below. In general, there is also a CP violating effective $ttgg$ vertex, the fourth diagram.
A convenient way to calculate the $CP$ asymmetry is to consider the process as in Figure~\ref{fig:fact} in the parton CM frame and use a mixed method of helicity amplitudes and traces of Dirac matrices as we described in Ref.~\cite{Valencia:2005cx}.  
The top-quark pair production by the four diagrams in Figure~\ref{fig:prod} is represented by $\Gamma_P$ in Figure~\ref{fig:fact}. The $t$ and $\bar{t}$ decays into $bW$ are represented by $\Gamma_{D,\bar{D}}$. We will consider two cases: first, we treat the $W$ as a final state, an approximation useful to describe hadronic $W$ decays where no correlations involving the decay products of the $W$ are observed; and second, we allow the $W$ to decay into $\ell \nu$ with a standard model vertex. 
The amplitude can then be written schematically as
\begin{equation}
{\cal M} =- \frac{\bar{u}_b\Gamma_D (\slashchar{p}_t+m_t)\Gamma_P 
(-\slashchar{p}_{\bar{t}}+m_t) \Gamma_{\bar{D}} v_{\bar{b}}}{(p_t^2-m_t^2)(p_{\bar{t}}^2-m_t^2)}.
\end{equation} 
We now split the production and decay processes using helicity amplitudes and replace the numerator of the top-quark (and anti-top-quark) propagator with a sum over polarizations. We work within the narrow-width approximation for the $t$ and $\bar{t}$ decays; and, therefore, these polarization sums refer to  on-shell $t\bar{t}$ states. Notice, however, that this procedure preserves the full spin correlations. As it turns out, the $CP$ odd observable arises from the interference of amplitudes in which the intermediate states have different helicities. Since the $b$ and the $\bar{b}$ polarizations are not observable, we sum over them immediately after squaring the amplitude.  Similarly, we sum over the $W$ polarization for the case of $W$ final states or over the $\ell$ and $\nu$ polarizations for the case when the $W$ decays leptonically. We thus write
\begin{eqnarray}
\left| {\cal M}\right|^2 &=& \left(\frac{\pi}{m_t\Gamma_t}\right)^2\delta(p_t^2-m_t^2)\delta(p_{\bar{t}}^2-m_t^2)\sum_{\lambda,\lambda',\sigma,\sigma'} {\cal T}_t(\lambda',\lambda) {\cal T}_{\bar t}(\sigma,\sigma') {\cal T}_{P}(\lambda,\sigma,\sigma',\lambda')
\label{msqb}
\end{eqnarray}
where we have defined the helicity factors
\begin{eqnarray}
 {\cal T}_t(\lambda',\lambda) &\equiv&
\left( \bar{u}_{t\lambda'}\gamma^0\Gamma^\dagger_{D}\gamma^0 \slashchar{p}_b \Gamma_{D}u_{t\lambda}\right)  \nonumber \\
{\cal T}_{\bar t}(\sigma,\sigma') &\equiv&
\left( \bar{v}_{\bar{t}\sigma} \Gamma_{\bar{D}} \slashchar{p}_{\bar{b}} \gamma^0\Gamma^\dagger_{\bar{D}}\gamma^0 v_{\bar{t}\sigma'} \right) \nonumber \\
{\cal T}_{P}(\lambda,\sigma,\sigma',\lambda') &\equiv&
 \left(
\bar{u}_{t\lambda}\Gamma_Pv_{\bar{t}\sigma}\ \ 
\bar{v}_{\bar{t}\sigma'}\gamma^0\Gamma^\dagger_P\gamma^0 u_{t\lambda'} \right)
\label{msqfacs}
\end{eqnarray}
To proceed, we consider several cases separately in what follows.

\section{CP Violation in the Production Vertex}

We first study CP violation in the production vertex, taking the decay vertices to proceed as in the standard model. $CP$ violation will be due to an effective dipole moment anomalous coupling of the top-quark defined via the Lagrangian
\begin{eqnarray}
{\cal L}_{cdm}&=&-ig_s\frac{\tilde{d}}{2}\bar{t}\sigma_{\mu\nu}\gamma_5\,G^{\mu\nu}\, t  
\end{eqnarray}
where $g_s$ is the strong coupling constant and $G^{\mu\nu}$ is the usual field strength tensor. This Lagrangian modifies the standard model top-quark couplings to gluons as follows (for incoming gluons that carry momentum $q$)
\begin{eqnarray}
gt\bar{t} &\to& -ig_s\frac{\lambda_a}{2}\left(\gamma_\mu+\tilde{d}\, \sigma_{\mu\nu}q^\nu\gamma_5\right) \nonumber \\
ggt\bar{t} &\to& i\, \pi \, \alpha_s\, [\lambda^b,\lambda^c]\, \tilde{d}\, \sigma_{\mu\nu}\gamma_5.
\label{effver}
\end{eqnarray}

The production factor becomes, summing over the gluon helicity $\lambda_{1,2}$, 
\begin{eqnarray}
{\cal T}_{P}(\lambda,\sigma,\sigma',\lambda') &=&
\frac{1}{4} \, \frac{g_s^4C_{ij}}{64}\sum_{\lambda_1,\lambda_2} {\cal M}_{Pi}(\lambda_1,\lambda_2,\lambda,\sigma)
{\cal M}_{Pj}^\star(\lambda_1,\lambda_2,\lambda',\sigma')
\label{prodfac}
\end{eqnarray}
where $M_{Pi}$ represents the helicity amplitudes for $gg\to t \bar{t}$ from channel $i=s,t,u$,  the factor $1/4$ accounts for the average over gluon helicities, and  $C_{ij}/64$ accounts for the color factor. Given the color structure of the effective vertices in Eq.~\ref{effver}, these color factors are the usual ones for $gg\to t\bar{t}$, namely:\footnote{The signs of these are defined by writting the amplitude as ${\cal M}={\cal M}_s[T_a,T_b]+{\cal M}_t T_a T_b + {\cal M}_u T_b T_a$.}  
\begin{equation}
C_{ss}=12, \, C_{st}=6, \, C_{su}=-6, \, C_{tt}=C_{uu}=\frac{16}{3},\, C_{tu} = -\frac{2}{3}.
\end{equation}
The new, seagull diagram has the same color structure as the $s$ channel diagram; hence, its associated color factors are the same as the corresponding ones for the $s$-channel amplitude. For our calculations we treat it as part of the $s$-channel amplitude. 
The helicity amplitudes corresponding to the production process are standard and can be found in the literature. For example, our amplitudes (including the anomalous top-quark coupling) agree with those in Ref.~\cite{Choi:1997ie}. 
As it turns out, this way of splitting the calculation simplifies it sufficiently that no explicit helicity amplitudes are needed.

For the decay factors we consider several cases.

\subsection{$W^\pm$ final states}

We begin by considering the case where the  $W$ is treated as a final state. We thus assume that its momentum can be reconstructed but not its polarization. That is, no angular correlations involving the $W$ decay products are measured. To consider this case 
we use the standard model vertex
\begin{equation}
\Gamma_D = \frac{g}{2\sqrt{2}}\gamma_\mu (1-\gamma_5) \epsilon^{\star\mu}_{W^+},
\end{equation}
the corresponding SM vertex $\Gamma_{\bar{D}}$ and 
sum over the $W^\pm$ polarization to obtain, 
\begin{eqnarray}
 {\cal T}_t(\lambda',\lambda) &=& \frac{g^2}{4}\bar{u}_{t\lambda'}\left(-m_t\left(1-\frac{m_t^2}{M_W^2}\right)+
\slashchar{p}_b\left(2-\frac{m_t^2}{M_W^2}\right)\right)
(1-\gamma_5) u_{t\lambda}
\nonumber \\
 {\cal T}_{\bar t}(\sigma,\sigma') &=& \frac{g^2}{4}
 \bar{v}_{\bar{t}\sigma}\left(m_t\left(1-\frac{m_t^2}{M_W^2}\right)+
\slashchar{p}_{\bar{b}}\left(2-\frac{m_t^2}{M_W^2}\right)\right)
(1-\gamma_5) v_{\bar{t}\sigma'}
\label{decfw}
\end{eqnarray}

In these two factors 
only the terms proportional to $\slashchar{p}_b$ and $\slashchar{p}_{\bar{b}}$ contribute to the $T$-odd correlations.

Combining all the different factors, we arrive at the final result. When we use the expressions in Eq.~\ref{decfw} in combination with the production factors of Eq.~\ref{prodfac}, it is possible to turn the sum over $t$ and $\bar t$ polarizations back into traces and compute the trace directly. It is also possible to obtain our results by using  explicit helicity amplitudes to sum over the intermediate states. We have checked our results by computing them with both methods.

Interestingly, we find that the $T$-odd correlations generated in this case are truly $CP$-odd and that they can be expressed in a factorized form: as a product of form factors that depend only on the $gg\to t\bar{t}$ production kinematic quantities $s,t,u$, and certain triple product correlations involving the momenta in the $t$ ($\bar{t})$ decay chains as well as the beam direction $p_1-p_2$. Moreover, the asymmetries are quadratic in this beam direction, ensuring their independence from the choice of $p_1$ between identical particles in the initial state. 

We express our results  for the triple products with a generic structure involving  the parton four-momentum sum and difference $P\equiv p_1+p_2$ and $q \equiv p_1-p_2$; the top anti-top momenta; and one momentum vector $p_D$ and $p_{\bar{D}}$ from the $t$ and $\bar{t}$ decay products respectively. From the invariant matrix element squared, we show {\bf only} those terms that lead to triple-product correlations. All such terms arise from the interference between the standard model amplitude and the $CP$ violating anomalous couplings\footnote{Here we use the Levi-Civita tensor contracted with four vectors $\epsilon(a,b,c,d) \equiv \epsilon_{\mu \nu \alpha \beta} a^\mu b^\nu c^\alpha d^\beta$ with the sign convention $\epsilon_{0123}=1$.}:
\begin{eqnarray}
\left| {\cal M}\right|_{CP}^2  &=&  \, C_1(s,t,u) \, q \cdot(p_{\bar{t}}-p_t) \,\epsilon(p_t,p_{\bar{t}},p_D,p_{\bar{D}})  \nonumber \\
&+& C_2(s,t,u) \,
\left(P \cdot p_t \, \epsilon(p_D,p_{\bar{D}},p_{\bar{t}},q) 
+P \cdot p_{\bar{t}} \, \epsilon(p_D,p_{\bar{D}},p_t,q) \right)
\nonumber \\
&+ & C_3(s,t,u)\, \left( P \cdot p_D \,
\epsilon(p_{\bar{D}},p_t,p_{\bar{t}},q)+P \cdot p_{\bar{D}} \,
\epsilon(p_D,p_t,p_{\bar{t}},q)\right)
\label{formfactors}
\end{eqnarray}
This form exhibits explicitly the symmetry between $t$ and $\bar t$ momenta, but it is also possible to simplify it further. For example,  
the factor in front of $C_1$ is just $ q \cdot(p_{\bar{t}}-p_t)= t-u$; and  similarly the factor multiplying $C_2$ can be written as $s \, \epsilon(p_D,p_{\bar D},P,q)/2$. The three form factors that appear in Eq.~\ref{formfactors} are independent as we have verified both with the use of all relevant Schouten identities \cite{schouten} and by explicitly constructing them in the parton center of mass frame. Using Schouten identities such as the one in the appendix, it is possible to rewrite them in different ways. 
For the case discussed in this subsection, the decay momenta entering Eq.~\ref{formfactors} are
\begin{equation}
p_D \to p_b, \,\, p_{\bar{D}} \to p_{\bar{b}}.
\end{equation}

There are $s,t,u$ channel contributions to the correlations, and we  present results separately for three different cases. All the form factors  include the overall factor
\begin{equation}
K_{bb} \equiv (\pi^2\alpha_s^2g^4 )\, \left(2-\frac{m_t^2}{M_W^2}\right)^2\, \left(\frac{\pi}{m_t\Gamma_t}\right)^2\, 
\delta(p_t^2-m_t^2)\, \delta(p_{\bar{t}}^2-m_t^2).
\end{equation}

The contribution from the $s$ channel amplitude squared is 
\begin{eqnarray}
C^s_1(s,t,u) = C^s_2(s,t,u) = C^s_3(s,t,u)  = \frac{3}{2} \, {\tilde d}\,  K_{bb}\, 
m_t\frac{(t-u)}{s^2}.
\label{s2cdm}
\end{eqnarray}
Notice that both the form factors in Eq.~\ref{s2cdm} and the correlations they multiply in Eq.~\ref{formfactors} are odd under the interchange of $p_1 \leftrightarrow p_2$. The combined effect being even under this interchange will not vanish after convolution with the parton distribution functions. 

The second contribution is that of the $t$ and $u$ channels. Adding up their squared amplitudes as well as the interference between them, we obtain
\begin{eqnarray}
C^{tu}_1(s,t,u) &=& \frac{1}{48}\, {\tilde d} \, K_{bb}\, \frac{m_t}{s^2
(t-m_t^2)^2(u-m_t^2)^2} \left[ 9 (t-u)^5 -2(5s-36m_t^2) s(t-u)^3\right. \nonumber \\
&&\left. +s^2 (s^2-22 s m_t^2 +144m_t^4)(t-u)+\frac{14m_t^2 s^4(s+8m_t^2)}{(t-u)}\right] \nonumber \\
C^{tu}_2(s,t,u) &=& \frac{1}{48}\, {\tilde d}\, K_{bb}\,  \frac{ m_t}{s^2(t-m_t^2)^2(u-m_t^2)^2}\left[9 (t-u)^5 -2  (5s-9{m_t^2})s(t-u)^3\right. \nonumber \\
&& \left. +s^2(s^2+46s{m_t^2}) (t-u) \right] \nonumber \\
C^{tu}_3(s,t,u) &=& C^{tu}_2(s,t,u).
\label{tu2cdm}
\end{eqnarray}
Once again notice that this contribution is even under the interchange of $p_1 \leftrightarrow p_2$ and will not vanish after convolution with the parton distribution functions. The last term for $C_1$ appears to have a factor of $(t-u)$ in the denominator, but notice that this is just an artifact of the notation in Eq.~\ref{formfactors}.

Finally, we compute the interference between the $s$ channel amplitude and the amplitudes for the $t$ and $u$ channels. We find
\begin{eqnarray}
C^{tu-s}_1(s,t,u) &=& -\frac{3}{4}\, {\tilde d}\, K_{bb}\, \frac{ m_t(t-u)}{s^2 (t-m_t^2)(u-m_t^2)}\left(-4sm_t^2+s^2-(t-u)^2\right) \nonumber \\
C^{tu-s}_2(s,t,u) &=&  -3 \, {\tilde d}\, K_{bb}\, m_t\frac{t-u}{s^2} \nonumber \\
C^{tu-s}_3(s,t,u) &=& C^{tu-s}_2(s,t,u).
\label{ts2cdm}
\end{eqnarray}
The form factors that appear in Eq.~\ref{formfactors} are thus the sum of the three contributions:
\begin{equation}
C_i(s,t,u)=C^s_i(s,t,u)+C^{tu}_i(s,t,u)+C^{tu-s}_i(s,t,u),
\end{equation}
for $i=1,2,3$. 

\subsection{Leptonic $W$ decay} 

Instead of summing over the $W$ polarization, we now allow it to decay leptonically with a standard model vertex. Using the narrow width approximation for the $W^\pm$,  Eq.~\ref{msqb} is trivially modified into
\begin{eqnarray}
\left| {\cal M}\right|^2 &=& \left(\frac{\pi}{m_t\Gamma_t}\right)^2\,  \left(\frac{\pi}{M_W\Gamma_W}\right)^2\delta(p_t^2-m_t^2)\delta(p_{\bar{t}}^2-m_t^2)\delta(p_{W^-}^2-M_W^2)\delta(p_{W^+}^2-M_W^2)
\nonumber \\
&\times& \sum_{\lambda,\lambda',\sigma,\sigma'} {\cal T}_t(\lambda',\lambda) {\cal T}_{\bar t}(\sigma,\sigma') {\cal T}_{P}(\lambda,\sigma,\sigma',\lambda')
\label{msqbp}
\end{eqnarray}
The decay vertex is given by
\begin{equation}
\Gamma_D = \left(\frac{g}{2\sqrt{2}}\right)^2\gamma_\mu (1-\gamma_5)\,
\left(-g^{\mu\alpha}+\frac{p_{W^+}^\mu p_{W^+}^\alpha}{M_W^2}\right)\, \bar{u}_\nu\gamma_\alpha  (1-\gamma_5) v_{\ell^+} ,
\end{equation}
(the index $\nu$ now denotes the neutrino) 
and a corresponding SM vertex $\Gamma_{\bar{D}}$. The helicity factors for the decay then become, after summing over the spin of both leptons, 
\begin{eqnarray}
 {\cal T}_t(\lambda',\lambda) &=& g^4 p_b\cdot p_\nu \, \bar{u}_{t\lambda'} \slashchar{p}_{\ell^+}(1-\gamma_5) u_{t\lambda}
\nonumber \\
 {\cal T}_{\bar t}(\sigma,\sigma') &=& g^4 p_{\bar{b}}\cdot p_{\bar{\nu}} \, 
 \bar{v}_{\bar{t}\sigma} \slashchar{p}_{\ell^-} (1-\gamma_5) v_{\bar{t}\sigma'}
\label{decfl}
\end{eqnarray}
It should be obvious by comparing Eq.~\ref{decfl} and Eq.~\ref{decfw} that, apart from overall constants, the asymmetries in this case can be obtained from the previous ones by replacing the $b$ momentum with the lepton momentum. So in Eq.~\ref{formfactors} we now have
\begin{equation}
p_D \to p_{\ell^+}, \,\, p_{\bar{D}} \to p_{\ell^-}
\end{equation}
The overall constant is now
\begin{eqnarray}
K_{\ell\ell} &\equiv & 16\, (\pi^2\alpha_s^2g^8 )\, \left(p_b\cdot p_\nu\right)\left( p_{\bar{b}}\cdot p_{\bar{\nu}} \right)\, \left(\frac{\pi}{m_t\Gamma_t}\right)^2\left(\frac{\pi}{M_W\Gamma_W}\right)^2 \nonumber \\
&\times&  \delta(p_t^2-m_t^2)\delta(p_{\bar{t}}^2-m_t^2) 
\delta(p_{W^+}^2-M_W^2)  \delta(p_{W^-}^2-M_W^2);
\end{eqnarray}
and the form factors are the same as those of Eqs.~\ref{s2cdm},~\ref{tu2cdm},~and~\ref{ts2cdm} with the replacement $K_{bb}\to K_{\ell\ell}$.

\subsection{One $W$ decays into leptons}

Finally, we consider the mixed case with final state  $b \ell^+ \nu W^-$ or $\bar{b} \ell^- \bar{\nu} W^+$. The final state is no longer a $CP$ eigenstate so the corresponding triple products for the two subcases are not $CP$-odd. $CP$ odd correlations can be constructed by adding the two possibilities. 
For the $b \ell^+ \nu W^-$ final state, the momenta in Eq.~\ref{formfactors} are 
\begin{equation}
p_D \to p_{\ell^+}, \,\, p_{\bar{D}} \to p_{\bar{b}},
\end{equation}
and the overall constant is now
\begin{eqnarray}
K_{\ell b} &\equiv & 4\, (\pi^2\alpha_s^2g^6 )\, \left(p_b\cdot p_\nu\right)\,
\left(2-\frac{m_t^2}{M_W^2}\right)\,
\left(\frac{\pi}{m_t\Gamma_t}\right)^2\left(\frac{\pi}{M_W\Gamma_W}\right) \nonumber \\
&\times&  \delta(p_t^2-m_t^2)\delta(p_{\bar{t}}^2-m_t^2) 
\delta(p_{W^+}^2-M_W^2). 
\end{eqnarray}
For the $\bar{b} \ell^- \bar{\nu} W^+$ final state, 
\begin{equation}
p_D \to p_{b}, \,\, p_{\bar{D}} \to p_{\ell^-};
\end{equation}
and the overall constant is
\begin{eqnarray}
K_{b\ell} &\equiv & 4\, (\pi^2\alpha_s^2g^6 )\, \left(p_{\bar{b}} \cdot 
p_{\bar{\nu}} \right)\, 
\left(2-\frac{m_t^2}{M_W^2}\right)\, 
\left(\frac{\pi}{m_t\Gamma_t}\right)^2\left(\frac{\pi}{M_W\Gamma_W}\right) \nonumber \\
&\times&  \delta(p_t^2-m_t^2)\delta(p_{\bar{t}}^2-m_t^2) 
 \delta(p_{W^-}^2-M_W^2).
\end{eqnarray}

\section{CP violation in the decay vertex}

We write the most general $tbW^+$ and $\bar{t} \bar{b} W^-$ vertices  as \cite{Bernreuther:1992be} (with complex form factors to allow for CP violation),
\begin{eqnarray}
\Gamma^\mu_{Wtb} &=& 
-\frac{g}{\sqrt{2}} \, V_{tb}^\star \,\bar{u}(p_b) \left[ \gamma_\mu (f_1^L P_L+f_1^R P_R)-
i  \sigma^{\mu\nu} (p_t-p_b)_\nu (f_2^L P_L+f_2^R P_R)\right) u(p_t), \nonumber \\
\bar\Gamma^\mu_{Wtb} &=&
-\frac{g}{\sqrt{2}} \, V_{tb} \,\bar{v}(p_{\bar{t}}) \left[ \gamma_\mu ({\bar f}_1^L P_L+{\bar f}_1^R P_R)-
i  \sigma^{\mu\nu} (p_{\bar t}-p_{\bar b})_\nu (\bar{f}_2^L P_L+\bar{f}_2^R P_R)\right) v(p_{\bar b}),
\end{eqnarray}
and for the remaining of the paper we will take $V_{tb}\equiv 1$. These vertices can be derived from a dimension five effective Lagrangian as in Ref.~\cite{otheranom}, and no seagulls  that contribute to the $T$-odd asymmetries we study are present. 

At tree-level within the SM the form factors $f_1^L={\bar f}_1^L=1$, while the other ones vanish. Under the assumption that the new CP violating interactions are smaller than the standard model interactions, we are only interested in those terms that can interfere with the SM and are therefore linear in the anomalous couplings. It is easy to see that only the term $f_2^R$ (${\bar f}_2^L$) generates triple product correlations via interference with the SM. To obtain signals that are only linear in new physics, we thus take
\begin{eqnarray}
f_1^L&=&{\bar f}_1^L\, =\, 1,  \nonumber \\
f_2^R \, =\, f e^{i(\phi_f+\delta_f)}, && 
{\bar f}_2^L \, =\,  f e^{i(-\phi_f+\delta_f)}.
\label{cpdecay}
\end{eqnarray}
We have introduced two types of phases: a CP-odd phase $\phi_f$, which can be introduced directly at the Lagrangian level; and a CP-even absorptive phase $\delta_f$. The latter arises from absorptive contributions beyond tree level and is the same for $t$ and ${\bar t}$ decay.  

At the top-quark decay level, with a polarized top-quark (anti-top) and with the $W$-boson decaying leptonically (but summing over the $b$-quark and lepton spin), the vertices in Eq.~\ref{cpdecay}, generate  T-odd triple products of the form
\begin{eqnarray}
d\Gamma(t\to b W^+) &\sim & f\sin(\delta_f+\phi_f)\epsilon(p_t,p_b,p_{\ell^+},s_t) \,+\, \cdots \nonumber \\
d\Gamma({\bar t}\to {\bar b} W^-) &\sim & f\sin(\delta_f-\phi_f)\epsilon(p_{\bar t},p_{\bar b},p_{\ell^-},s_{\bar t}) \,+\, \cdots
\label{decaybasic}
\end{eqnarray}
When the top-quark (anti-top) decay is connected with the gluon fusion production of $t{\bar t}$, these correlations will give rise to ones in which the top-quark (anti-top) spin is analyzed by a four vector from the production process or by one from the decay of the anti-top (top) quark. 

The helicity factors for the decay of $t$ and $\bar t$ of Eqs.~\ref{decfw},~\ref{decfl} become, for $W$ final states:
\begin{eqnarray}
 {\cal T}_t(\lambda',\lambda) &=& \frac{g^2}{4}\bar{u}_{t\lambda'}\slashchar{p}_b\left(\left(2-\frac{m_t^2}{M_W^2}\right)
(1-\gamma_5) +2i f m_t \sin(\phi_f+\delta_f) \right) u_{t\lambda}
\nonumber \\
 {\cal T}_{\bar t}(\sigma,\sigma') &=& \frac{g^2}{4}
 \bar{v}_{\bar{t}\sigma}
\slashchar{p}_{\bar{b}}\left( \left(2-\frac{m_t^2}{M_W^2}\right)
(1-\gamma_5)  +2i f m_t \sin(\phi_f-\delta_f) \right) v_{\bar{t}\sigma'}
\label{decfwv}
\end{eqnarray}
and for leptonic final states
\begin{eqnarray}
 {\cal T}_t(\lambda',\lambda) &=& g^4 p_b\cdot p_\nu \, \bar{u}_{t\lambda'} \slashchar{p}_{\ell^+}\left( (1-\gamma_5)  -2if \sin(\phi_f+\delta_f) \slashchar{p}_{\nu} \right) u_{t\lambda}
\nonumber \\
 {\cal T}_{\bar t}(\sigma,\sigma') &=& g^4 p_{\bar{b}}\cdot p_{\bar{\nu}} \, 
 \bar{v}_{\bar{t}\sigma} \slashchar{p}_{\ell^-}\left( (1-\gamma_5) +2if\sin(\phi_f-\delta_f) \slashchar{p}_{\bar \nu} \right)  v_{\bar{t}\sigma'}
\label{decflv}
\end{eqnarray}
In Eqs.~\ref{decfwv},~\ref{decflv} we have omitted terms that do not contribute to the triple products.

In all cases we will write the triple product correlations in the form
\begin{eqnarray}
|{\cal M}|^2_{T} &=& \, 
f\sin(\phi_f+\delta_f)\, \epsilon(p_t,p_{ b},p_{\ell^+},Q_{t}) +
 f\sin(\phi_f-\delta_f) \,\epsilon(p_{\bar t},p_{ \bar{b}},p_{\ell^-},Q_{\bar{t}}) 
 \label{asymcpdec}
\end{eqnarray} 
This form occurs naturally in the calculation: the first term arising from $CP$ violation in polarized top-quark decay with the top-spin being analyzed by the four vector $Q_t$, a linear combination of $p_{\bar t}$, $p_{\ell^-}$ and $q$. Correspondingly, the second term arises from the anti-top quark decay. Not all the terms in this expression violate $CP$ as is manifest by the presence of the strong phase $\delta$. 

In the case of $CP$ violation in the decay vertex, the initial $T$-odd spin correlations of Eq.~\ref{decaybasic} occur only when the $W$ decays as well. Thus, unlike the previous section, we can only consider two cases: when both $W^\pm$ decay leptonically and when at least one of them does. The two $W$ final-state case does not reveal $T$-odd correlations originating in the top-quark decay vertices.

Unlike two of the cases studied in the previous section, the $T$-odd  correlations in Eq.~\ref{asymcpdec} are not $CP$ odd. They are generated both by $CP$-violating phases and by $CP$-conserving absorptive phases. To construct truly $CP$-odd observables, it is necessary to compare the distributions in top-quark decay with those in anti-top quark decay. One way to do that is to notice that parts of Eq.~\ref{asymcpdec} can be written in the form of a truly $CP$-odd correlation $\epsilon(p_t,p_{\bar t},p_{\ell^+},p_{\ell^-})$ with the aid of Schouten identities as those in Eq.~\ref{moreschouten} in the appendix.

\subsection{Leptonic $W$ decay}

Once again we provide separate expressions for the contributions of $s$-channel amplitude squared; $t$ and $u$-channel amplitudes squared; and interference between the $s$ channel amplitude and those from $t$ and $u$ channels. For the contribution from the $s$-channel amplitude squared, we obtain:

\begin{eqnarray}
  Q_{t} &=& -K_{\ell \ell} \, \frac{3 m_t}{2s^2} \left\{((t-u)^2-s^2)p_{\ell^-} +
 2(s p_{\ell^-}\cdot(p_t+p_{\bar t}) -(t-u)p_{\ell^-} \cdot q)p_{\bar t}\right.
 \nonumber \\
& +&\left. 2((t-u)p_{\ell^-}\cdot (p_t+p_{\bar t}) -s p_{\ell^-} \cdot q) q \right\}
\nonumber \\
   Q_{\bar t} &=&-K_{\ell \ell} \, \frac{3 m_t}{2s^2} \left\{ ((t-u)^2-s^2)p_{\ell^+} +
 2(s p_{\ell^+}\cdot(p_t+p_{\bar t}) +(t-u)p_{\ell^+} \cdot q)p_{ t}
 \right. \nonumber \\
& -&\left. 2((t-u)p_{\ell^+}\cdot (p_t+p_{\bar t}) +s p_{\ell^+} \cdot q) q
\right\}
\label{cpdecs}
\end{eqnarray}

For the $t$ and $u$-channels squared (plus their interference), we find:

\begin{eqnarray}
Q_t &=&-K_{\ell \ell} \,  \frac{m_t}{3 s^2 (s^2-(t-u)^2)^2}\left\{ 
\left( 16(7s^4+9(t-u)^2s^2)m_t^4 \right. \right. \nonumber \\
&-& \left. 
4(7s^5+11(t-u)^2s^3-18(t-u)^4s)m_t^2-(s^2-9(t-u)^2)(s^2-(t-u)^2)^2\right)p_{\ell^-} 
 \nonumber \\
&+& 2\left(s(9(t-u)^4+2(9m_t^2-5s)s(t-u)^2+s^3(14m_t^2+s))p_{\ell^-}\cdot p_t     
+ s(8s^4+(t-u)^2s^2 -9(t-u)^4 \right. \nonumber \\
&-& \left. 6m_t^2(7s^3 
+9(t-u)^2s))p_{\ell^-}\cdot p_{\bar t} 
+ (t-u)(s^2-9(t-u)^2)(4sm_t^2-s^2+(t-u)^2)p_{\ell^-}\cdot q
 \right) p_{\bar t} \nonumber \\
 &-& 2 \left( (t-u)(s^2-9(t-u)^2)(4sm_t^2-s^2+(t-u)^2)p_{\ell^-}\cdot p_t
 +(t-u)(-10s^4+19(t-u)^2s^2 \right. \nonumber \\
 &-& 9(t-u)^4 + 4m_t^2(s^3-9s(t-u)^2))p_{\ell^-}\cdot p_{\bar t} 
 + s(s^4-10(t-u)^2s^2 \nonumber \\
 &+& \left. \left. 9(t-u)^4+2m_t^2 (7s^3+9(t-u)^2s))p_{\ell^-}\cdot q   \right) q\right\}
\nonumber \\
   Q_{\bar t} &=&- K_{\ell \ell} \, \frac{m_t}{3 s^2 (s^2-(t-u)^2)^2}\left\{ 
\left( 16(7s^4+9(t-u)^2s^2)m_t^4 \right. \right. \nonumber \\
&-& \left. 
4(7s^5+11(t-u)^2s^3-18(t-u)^4s)m_t^2-(s^2-9(t-u)^2)(s^2-(t-u)^2)^2\right) p_{\ell^+}  \nonumber \\
   &-& 2\left(s(9(t-u)^4+2(9m_t^2-5s)s(t-u)^2+s^3(14m_t^2+s))p_{\ell^+}\cdot p_{\bar t}     
- s(8s^4+(t-u)^2s^2 -9(t-u)^4 \right. \nonumber \\
&-& \left. 6m_t^2(7s^3 
+9(t-u)^2s))p_{\ell^+}\cdot p_{ t} 
+ (t-u)(s^2-9(t-u)^2)(4sm_t^2-s^2+(t-u)^2)p_{\ell^+}\cdot q
 \right)  p_{ t} \nonumber \\
& +& 2 \left( -(t-u)(s^2-9(t-u)^2)(4sm_t^2-s^2+(t-u)^2)p_{\ell^+}\cdot p_{\bar t}
 +(t-u)(10s^4-19(t-u)^2s^2\right. \nonumber \\
 &+& 9(t-u)^4 - 4m_t^2(s^3-9s(t-u)^2))p_{\ell^+}\cdot p_{ t} 
 + s(s^4-10(t-u)^2s^2+9(t-u)^4 \nonumber \\
 &+&\left. \left.2m_t^2 (7s^3+9(t-u)^2s))p_{\ell^+}\cdot q   \right) q \right\}
 \label{cpdectu}
 \end{eqnarray}
 
Finally, for the interference between the $s$-channel amplitude and those from the $t$ and $u$ channels, we find

\begin{eqnarray}
Q_t &=& K_{\ell \ell} \, \frac{3 m_t}{s^2 (s^2-(t-u)^2)} \left\{ \left(s^4-2(s-2m_t^2)(t-u)^2s+(t-u)^4 \right)   p_{\ell^-} \right. \nonumber \\
& -& 2\left( s^3 p_{\ell^-}\cdot p_{\bar t} +(s^3-s(t-u)^2)p_{\ell^-}\cdot p_t +(t-u)(2sm_t^2-s^2+(t-u)^2)p_{\ell^-}\cdot q \right) p_{\bar t} \nonumber \\
&+& 2\left((t-u)(2sm_t^2-s^2+(t-u)^2)p_{\ell^-}\cdot p_t +(t-u)(2sm_t^2-2s^2+(t-u)^2)p_{\ell^-}\cdot p_{\bar t} \right. \nonumber \\
&+& \left.  \left. s(s^2-(t-u)^2)p_{\ell^-}\cdot q \right) q \right\}
  \nonumber \\
   Q_{\bar t} &=& K_{\ell \ell} \, \frac{3 m_t}{s^2 (s^2-(t-u)^2)} \left\{ \left(s^4-2(s-2m_t^2)(t-u)^2s+(t-u)^4 \right)   p_{\ell^+} \right. \nonumber \\
& -& 2\left( s^3 p_{\ell^+}\cdot p_{ t} +(s^3-s(t-u)^2)p_{\ell^+}\cdot 
p_{\bar t} -(t-u)(2sm_t^2-s^2+(t-u)^2)p_{\ell^+}\cdot q \right) p_{ t} \nonumber \\
&-& 2\left((t-u)(2sm_t^2-2s^2+(t-u)^2)p_{\ell^+}\cdot p_{t} +(t-u)(2sm_t^2-s^2+(t-u)^2)p_{\ell^+}\cdot p_{\bar t} \right. \nonumber \\
&-& \left.  \left. s(s^2-(t-u)^2)p_{\ell^+}\cdot q \right) q \right\}
\label{cpdects}
\end{eqnarray}

\subsection{Only one $W$ decays into leptons}

When the $W^+$ decays into leptons, only the first term of Eq.~\ref{asymcpdec} is present.  Formally $Q_{\bar t} = 0$ and $Q_t$ can be obtained simply from Eqs.~\ref{cpdecs},~\ref{cpdectu},~and~\ref{cpdects} with the replacements 
\begin{eqnarray}
p_{\ell^-} &\to & p_{\bar b} \nonumber \\
K_{\ell \ell} &\to&  K_{\ell b}.
\label{rep1}
\end{eqnarray}

When the $W^-$ decays into leptons, only the second term of Eq.~\ref{asymcpdec} is present.  Formally $Q_{t} = 0$ and $Q_{\bar t}$ can be obtained simply from Eqs.~\ref{cpdecs},~\ref{cpdectu},~and~\ref{cpdects} with the replacements 
\begin{eqnarray}
p_{\ell^+} &\to & p_{b} \nonumber \\
K_{\ell \ell} &\to & K_{b \ell}.
\label{rep2}
\end{eqnarray}

\section{Light $q{\bar q}$ annihilation}

The $q{\bar q}$ production mechanism can be treated in a similar manner. Ignoring $CP$ violation in the light quark couplings it is possible to obtain the corresponding results from the above formalism by adopting the notation 
\begin{equation}
\Gamma_P^{q{\bar q}} \equiv {\bar v}_q\gamma^\mu u_q \gamma_\mu
\end{equation}
and using the appropriate color/spin factor $g_s^4/18$. 

For $CP$ violation in the production vertex with $W^\pm$ final states we find 
\begin{eqnarray}
C^{q{\bar q}}_1(s,t,u)& =& -\frac{16}{9} \, {\tilde d}\,  K_{bb}\, 
m_t\left( \frac{(t-u)}{s^2}+4\frac{m_t^2}{s(t-u)}\right), \nonumber \\
C^{q{\bar q}}_2(s,t,u) &=& C^{q{\bar q}}_3(s,t,u) \,  = \, -\frac{16}{9} \, {\tilde d}\,  K_{bb}\, 
m_t\frac{(t-u)}{s^2}.
\label{s2cdmqq}
\end{eqnarray}
There is a term with an apparent factor of $(t-u)$ in the denominator, but recall that this is cancelled out by the normalization in the definition of these form factors, Eq.~\ref{formfactors}. The corresponding cases of leptonic final states or one leptonic and one $W$ final states are obtained from Eq.~\ref{s2cdmqq} with the same replacements discussed for the gluon fusion mechanism.

For $CP$ violation in the decay vertex and leptonic final states we obtain
\begin{eqnarray}
  Q^{q{\bar q}}_{t} &=& K_{\ell \ell} \, \frac{16 m_t}{9s^2} \left\{(4 s m_t^2+(t-u)^2-s^2)p_{\ell^-} +
 2(s p_{\ell^-}\cdot(p_t-p_{\bar t}) -(t-u)p_{\ell^-} \cdot q)p_{\bar t}\right.
 \nonumber \\
& +&\left. 2((t-u)p_{\ell^-}\cdot (p_t+p_{\bar t}) -s p_{\ell^-} \cdot q) q \right\}
\nonumber \\
   Q^{q{\bar q}}_{\bar t} &=&K_{\ell \ell} \, \frac{16 m_t}{9s^2} \left\{ (4sm_t^2+(t-u)^2-s^2)p_{\ell^+} -
 2(s p_{\ell^+}\cdot(p_t-p_{\bar t}) -(t-u)p_{\ell^+} \cdot q)p_{ t}
 \right. \nonumber \\
& -&\left. 2((t-u)p_{\ell^+}\cdot (p_t+p_{\bar t}) +s p_{\ell^+} \cdot q) q
\right\}.
\label{cpdecsqq}
\end{eqnarray}
For one $W$ and one leptonic final state the same replacements of Eqs.~\ref{rep1}~and~\ref{rep2}.

\section{Comparison with the literature}

The results we present in Eq.~\ref{formfactors}-Eq.~\ref{ts2cdm} are obtained from the manipulations implied by Eq.~\ref{msqfacs}. The three factors in Eq.~\ref{msqfacs} correspond to the decay density matrix for the top-quark, the decay density matrix for the anti-top-quark and the production density matrix for $gg\to t\bar{t}$ respectively. As such, these factors have been computed before for the case of anomalous top-quark coupling discussed here. For example, the production helicity amplitudes are explicitly given in Ref.~\cite{Choi:1997ie} and the production density matrix in the parton center of mass frame can be found in the Appendix of Ref.~\cite{Brandenburg:1992be}. The decay density matrix for the top-quark in its rest frame is also found in Ref.~\cite{Ma:1991ry}. These results are not sufficient to reproduce ours. For example if one starts from Eq. A4 of Ref.~\cite{Brandenburg:1992be}, the production density matrix in the parton center of mass frame, one also needs the corresponding decay density matrices in the same (parton center of mass) frame. We calculate the latter and find (for the case where the $W$ doesn't decay, for example), 
\begin{eqnarray}
{\cal T}_t(\lambda',\lambda)  &\sim & \frac{m_t^2-M_W^2}{E_1+m_t} +\frac{2m_t}{E_1+m_t}\vec\sigma_+\cdot\vec{p}_b-\frac{m_t^2-M_W^2+2E_bm_t}{(E_1+m_t)^2}\vec\sigma_+\cdot\vec{k}_+
\end{eqnarray}
where we have used the notation of Eq. A4 of Ref.~\cite{Brandenburg:1992be}, and there is a corresponding expression for the anti-top-quark decay. The terms proportional to $\sigma_+\cdot\vec{p}_b$ (and $\sigma_-\cdot\vec{p}_{\bar b}$ from anti-top decay) can be easily combined with Eq. A4 of Ref.~\cite{Brandenburg:1992be} to yield contributions to $C_1$ and $C_2$. With the terms proportional to $\sigma_+\cdot\vec{k}_+$ one needs to use the Schouten identity of our Appendix, Eq.~\ref{schou}, to obtain contributions to all three form factors $C_1,C_2,C_3$. When all this is done, we find  agreement with our result in Eqs.~\ref{formfactors}-\ref{ts2cdm} after expressing   it in the parton center of mass frame. The corresponding production density matrix for the derivation of Eq.~\ref{s2cdmqq} is Eq.~A3 of Ref.~\cite{Brandenburg:1992be}.

The first paper in Ref.~\cite{hadronanom}, D. Atwood {\it et. al.} performs a similar but simpler calculation. Instead of computing the density matrices, that reference computes the production of on-shell polarized top quarks and argues that the lepton momentum in the subsequent semi-leptonic top-quark decay acts as a spin analyser. 
This simpler calculation misses the contributions from off-diagonal entries in the spin density matrices, corresponding to interference between diagrams containing intermediate top-quarks with different helicities. Our result, therefore, disagrees with this reference.

\section{Conclusion}

We have revisited the question of $T$-odd triple product correlations in $t {\bar t}$ production and decay arising from anomalous top-quark couplings. We have illustrated a method of simplifying the calculation that allows us to obtain complete analytic expressions for the results. The main results of our paper are thus Eqs.~\ref{formfactors}-\ref{ts2cdm} and Eqs.~\ref{cpdecs}-\ref{cpdects}.

When $CP$ is violated in the production vertex, we obtain $T$-odd correlations that are truly $CP$-odd as well. In contrast, $CP$ violation in the decay vertex leads to $T$-odd correlations that can be faked by unitary ($CP$-conserving) phases. It is possible to turn these correlations into true $CP$-odd observables by comparing the $t$ and $\bar t$ decays. 

Our results fully incorporate the effect of all spin correlations and should be easy to implement in simulations that use the narrow width approximation for top-quark and $W$-boson propagators. 

The sensitivity of the LHC to the coupling $\tilde{d}$ has been studied before. For example, Ref.~\cite{Sjolin:2003ah} finds that ATLAS may achieve a $5\sigma$ sensitivity of $\tilde{d} < 26.3 \times 10^{-5}$~GeV$^{-1}$ (or $\tilde{d}/m_t < 0.046$) with $10~ fb^{-1}$ of data using both purely leptonic and one $W$ decaying into leptons final states, and with certain assumptions about other anomalous couplings. Similarly, Ref.~\cite{AguilarSaavedra:2007rs} finds that ATLAS may achieve a $2 \sigma$ sensitivity $-0.026 \leq f/M_W \leq 0.0312$.

\begin{acknowledgments}

This work was supported in part by DOE under contract number DE-FG02-01ER41155. 
G.V. thanks the Cavendish Laboratory at the University of Cambridge and CERN for their hospitality
while this work was completed. We thank Yili Wang for collaboration in the early stages of this work and F. del Aguila, J.~A.~Aguilar-Saavedra and D. Atwood for useful remarks.

\end{acknowledgments}

\appendix

\section{Identities}

Several identities involving the epsilon tensor (Schouten identities) were used. For $CP$ violation in the production vertex we found the following identity useful:
\begin{eqnarray}
&& \left ( p_b\cdot(p_1+p_2) 
 \epsilon(p_{\bar{b}},p_t,p_{\bar{t}},p_1-p_2)+p_{\bar{b}}\cdot(p_1+p_2) 
\epsilon(p_b,p_t,p_{\bar{t}},p_1-p_2)\right) = \nonumber \\ &&
(p_t-p_{\bar{t}})\cdot (p_1-p_2) \epsilon(p_t,p_{\bar{t}},p_b,p_{\bar{b}})
+ \nonumber \\ &&
(2m_t^2-s/2)\epsilon(p_b,p_{\bar{b}},p_t+p_{\bar{t}},p_1-p_2)+\nonumber \\  &&
(m_t^2-M_W^2)\epsilon(p_b+p_{\bar{b}},p_t,p_{\bar{t}},p_1-p_2).
\label{schou}
\end{eqnarray}
For $CP$ violation in the decay vertex the following identities are useful:
\begin{eqnarray}
P\cdot p_b \epsilon(p_t,p_{\bar t},p_{\ell^+},p_{\ell^-}) &=& 
\frac{s}{2}\left( \epsilon(p_b,p_{\bar t},p_{\ell^+},p_{\ell^-})+   \epsilon(p_t,p_b,p_{\ell^+},p_{\ell^-})\right) \nonumber \\
&+& \left(P\cdot p_{\ell^+}  \epsilon(p_t,p_{\bar t},p_b,p_{\ell^-})-P\cdot p_{\ell^-} \epsilon(p_t,p_b,p_{\ell^+},p_{\bar t})\right) \nonumber \\
P\cdot p_{\bar b} \epsilon(p_t,p_{\bar t},p_{\ell^+},p_{\ell^-}) &=& 
\frac{s}{2}\left( \epsilon(p_{\bar t},p_{\bar b},p_{\ell^-},p_{\ell^+})+   \epsilon(p_t,p_{\bar b},p_{\ell^+},p_{\ell^-})\right) \nonumber \\
&+& \left(P\cdot p_{\ell^-} \epsilon(p_{\bar t},p_t,p_{\bar b},p_{\ell^+})-P\cdot p_{\ell^+}  \epsilon(p_{\bar t},p_{\bar b},p_{\ell^-},p_t)\right)
\label{moreschouten}
\end{eqnarray}

\end{document}